\documentclass[aps,prl,superscriptaddress,amssymb,reprint]{revtex4-1}
\pdfoutput=1
\usepackage{amsmath, amsfonts}    
\usepackage{graphicx}   
\usepackage{verbatim}   
\usepackage{color}      
\usepackage{subfigure}  
\usepackage{hyperref}   
\usepackage{natbib}
\usepackage[USenglish]{babel}

\newcommand{\bra}[1]{\langle #1|}
\newcommand{\ket}[1]{\left|#1\right\rangle}

\begin{document}

\title{Manipulating a qubit through the backaction of sequential partial measurements \\ and real-time feedback }
\author{M.S.~Blok}
\email{These authors contributed equally to this work.}
\author{C.~Bonato}
\email{These authors contributed equally to this work.}
\affiliation{Kavli Institute of Nanoscience Delft, Delft University of Technology, P.O. Box 5046, 2600 GA Delft, The Netherlands}
\author{M.L.~Markham}
\author{D.J.~Twitchen}
\affiliation{Element Six Ltd, Kings Ride Park, Ascot, Berkshire SL5 8BP, UK.}
\author{V.V.~Dobrovitski}
\affiliation{Ames Laboratory and Iowa State University, Ames, Iowa 50011, USA.}
\author{R.~Hanson}
\email{r.hanson@tudelft.nl}
\affiliation{Kavli Institute of Nanoscience Delft, Delft University of Technology, P.O. Box 5046, 2600 GA Delft, The Netherlands}

\begin{abstract}
Quantum measurements not only extract information from a system but also alter its state. Although the outcome of the measurement is probabilistic, the backaction imparted on the measured system is accurately described by quantum theory~\cite{Guerlin_Nature_2007,Hatridge_Science_2013,Murch_Nature_2013}. Therefore, quantum measurements can be exploited for manipulating quantum systems without the need for control fields~\cite{Ashhab_PhysRevA_2010,Wiseman_NatureNV_2011}. We demonstrate measurement-only state manipulation on a nuclear spin qubit in diamond by adaptive partial measurements. We implement the partial measurement via tunable correlation with an electron ancilla qubit and subsequent ancilla readout~\cite{Brun_PhysRevA_2008,Groen_PRL_2013}. We vary the measurement strength to observe controlled wavefunction collapse and find post-selected quantum weak values~\cite{Brun_PhysRevA_2008,Groen_PRL_2013,Aharonov_PRL_1988,Pryde_PRL_2005,Dressel_ArXiv_2013}. By combining a novel quantum non-demolition readout on the ancilla with real-time adaption of the measurement strength we realize steering of the nuclear spin to a target state by measurements alone. Besides being of fundamental interest, adaptive measurements can improve metrology applications~\cite{Cappellaro_PhysRevA_2012,Higgins_Nature_2007} and are key to measurement-based quantum computing~\cite{Raussendorf_PRL_2001,Prevedel_Nature_2007}.
\end{abstract}

\maketitle

Measurements play a unique role in quantum mechanics and in quantum information processing. The backaction of a measurement can be used for state initialization~\cite{Robledo_Nature_2011,Riste_PRL_2012}, generation of entanglement between non-interacting systems~\cite{Chou_Nature_2005,Moehring_Nature_2007,Pfaff_NatPhys_2012,Riste_Nature_2013}, and for qubit error detection~\cite{Chiaverini_Nature_2004}. These measurement-based applications require either post-selection or real-time feedback, since the outcome of a measurement is inherently probabilistic. Recent experiments achieved quantum feedback control on a single quantum system~\cite{Riste_Nature_2013, Gillett_PRL_2010,Sayrin_Nature_2011,Vijay_Nature_2012} by performing coherent control operations conditioned on a measurement outcome.

Here, we realize real-time adaptive measurements and exploit these in a proof-of-principle demonstration of measurement-only quantum feedback. Our protocol makes use of partial measurements that balance the information gain and the measurement backaction by varying the measurement strength. We accurately control the measurement strength and the corresponding backaction in a two-qubit system by tuning the amount of (quantum) correlation between the system qubit and an ancilla qubit, followed by projective readout of the ancilla~\cite{Brun_PhysRevA_2008,Groen_PRL_2013}. In general, the backaction of sequential partial measurements leads to a random walk~\cite{Guerlin_Nature_2007,Hatridge_Science_2013,Murch_Nature_2013} but by incorporating feedback, multiple measurements can direct the trajectory of a qubit towards a desired state~\cite{Ashhab_PhysRevA_2010,Wiseman_NatureNV_2011}. Real-time adaptive measurements are a key ingredient for quantum protocols such as one-way quantum computing~\cite{Raussendorf_PRL_2001,Prevedel_Nature_2007} and Heisenberg-limited phase estimation~\cite{Cappellaro_PhysRevA_2012,Higgins_Nature_2007}.

We implement the adaptive partial measurements in a nitrogen vacancy (NV) center in synthetic diamond. We define the system qubit by the nuclear spin of the NV host nitrogen ($\ket{\downarrow}$: $m_I$=0, $\ket{\uparrow}$: $m_I$= -1), and the ancilla qubit by the NV electron spin ($\ket{0}$: $m_S$=0, $\ket{1}$: $m_S$=-1) (Fig.~1a). The ancilla is initialized and read out in a single shot with high fidelity using spin-selective optical transitions~\cite{Robledo_Nature_2011}. We perform single-qubit operations on the ancilla by applying microwave frequency pulses to an on-chip stripline.

\begin{figure*}
	\includegraphics{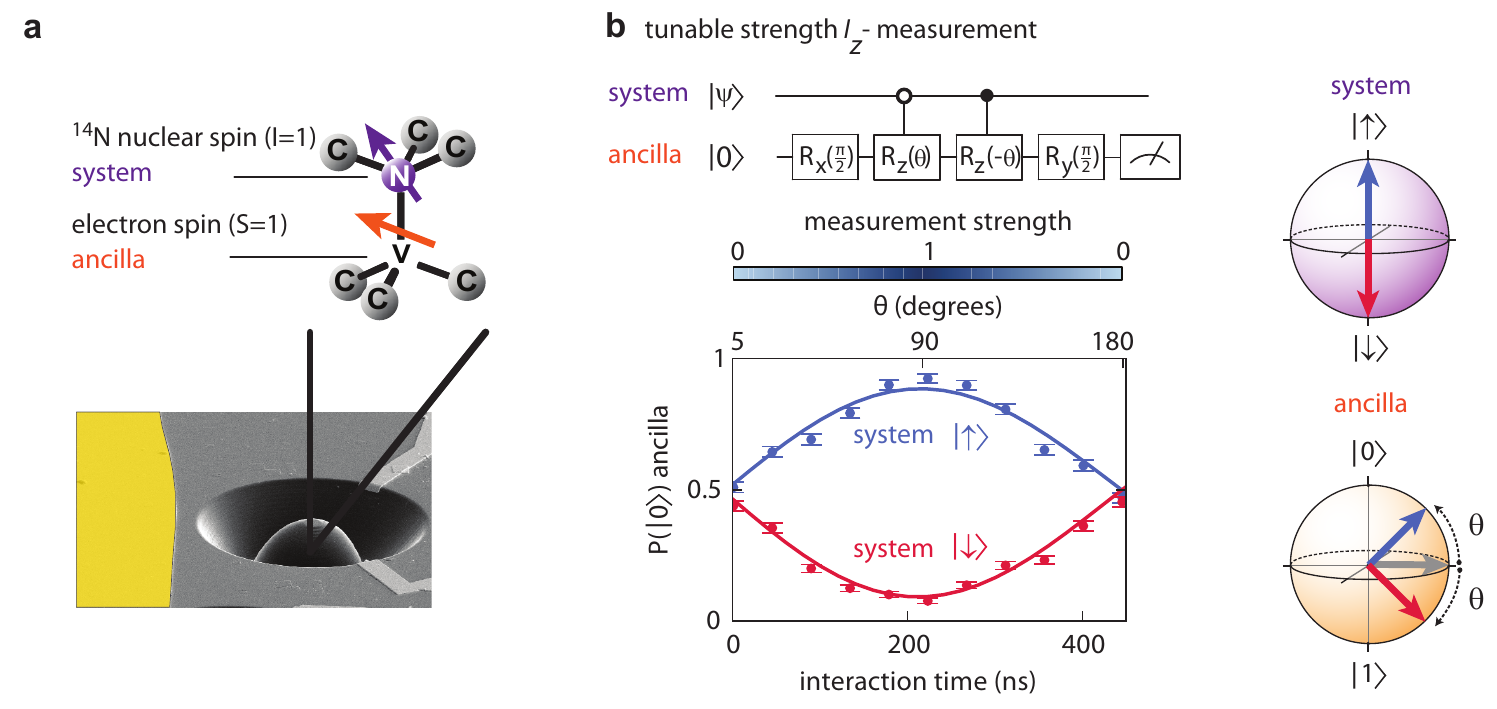}
	\caption{\label{fig1}Partial measurement of a spin qubit in diamond. (a) The NV center is a natural two-qubit system where the system qubit is defined by the $^{14}N$ nuclear spin and the ancilla qubit is defined by the electron spin. A solid-immersion-lens is deterministically fabricated on top of the selected NV center to increase the photon collection efficiency. Control fields for single qubit rotations are generated by applying a current to the gold stripline (yellow).  (b) A tunable strength measurement is implemented by a Ramsey-type gate on the ancilla. We plot the probability to measure the state $\ket{0}$  for the ancilla, as a function of interaction time τ, for two system input states $\ket{\downarrow}$ (red) and $\ket{\uparrow}$ (blue). The Bloch-spheres show the state of the system (purple) and ancilla (orange) after the entangling-gate for the different input states (red and blue vectors). The color bar represents the measurement strength, proportional to $\sin{\theta}$, where $\theta=\frac{A \tau}{2}$. Blue corresponds to a projective measurement and white to no measurement. Solid lines are a  fit to the function $y_0 + e^{-( \frac{\tau}{T_2^*})^2} \cos{(A \tau + \phi)} $. From the fit we find the weakest measurement we can perform, corresponding to $\theta = 5^{\circ}$. This is limited by free evolution of the ancilla during the pulses. }
\end{figure*}

We realize the variable-strength measurement by correlating the system qubit with the ancilla through a controlled-phase-type gate (Fig.~1b) that exploits the hyperfine interaction, which (neglecting small off-diagonal terms) has the form $\hat{H}_{hf}=A\hat{S}_{z}\hat{I}_{z}$ (with $A = 2 \pi \times 2.184 \pm 0.002$ MHz).  During free evolution, the ancilla qubit precession is conditional on the state of the system qubit. We choose the rotating frame such that the ancilla rotates clockwise (anti-clockwise) around the z-axis if the system qubit is in $\ket{\uparrow}$ ($\ket{\downarrow}$) and vary the interaction time $\tau$. For $\tau = 0$, there is no correlation between the ancilla and the system, while for $\tau = \frac{\pi}{A}$, corresponding to the rotation angle $\theta = 90^{\circ}$, the two are maximally correlated. A subsequent rotation and projective readout of the ancilla then implements a measurement of the system qubit, with a measurement strength that can be accurately tuned by controlling the interaction time $\tau$. 

We investigate the measurement-induced backaction by performing state tomography of the post-measurement state of the system (Fig.~2). First, we neglect the outcome of the partial measurement, which is mathematically equivalent to taking the trace over the state of the ancilla qubit. In this case the backaction is equivalent to pure dephasing as can be seen by a measured reduction of the length of the Bloch vector (Fig.~2b). Next, we condition the tomography on the ancilla measurement yielding state $\ket{0}$ (Fig.~2c). We observe that for a weak measurement $(\theta = 5^{\circ})$, the system is almost unaffected, while for increasing measurement strength it receives a stronger kick towards $\ket{\uparrow} $(Fig.~2c). Crucially, we find that the length of the Bloch vector is preserved in this process, which shows that the partial collapse is equivalent to a qubit rotation that is conditional on the measurement strength and outcome and on the initial state. By performing quantum process tomography, we find that both measurement processes agree well with the theoretical prediction (the process fidelities are 0.986 $\pm$ 0.004 and 0.94 $\pm$ 0.01 for the unconditional and conditional process, respectively).

\begin{figure}
	\includegraphics{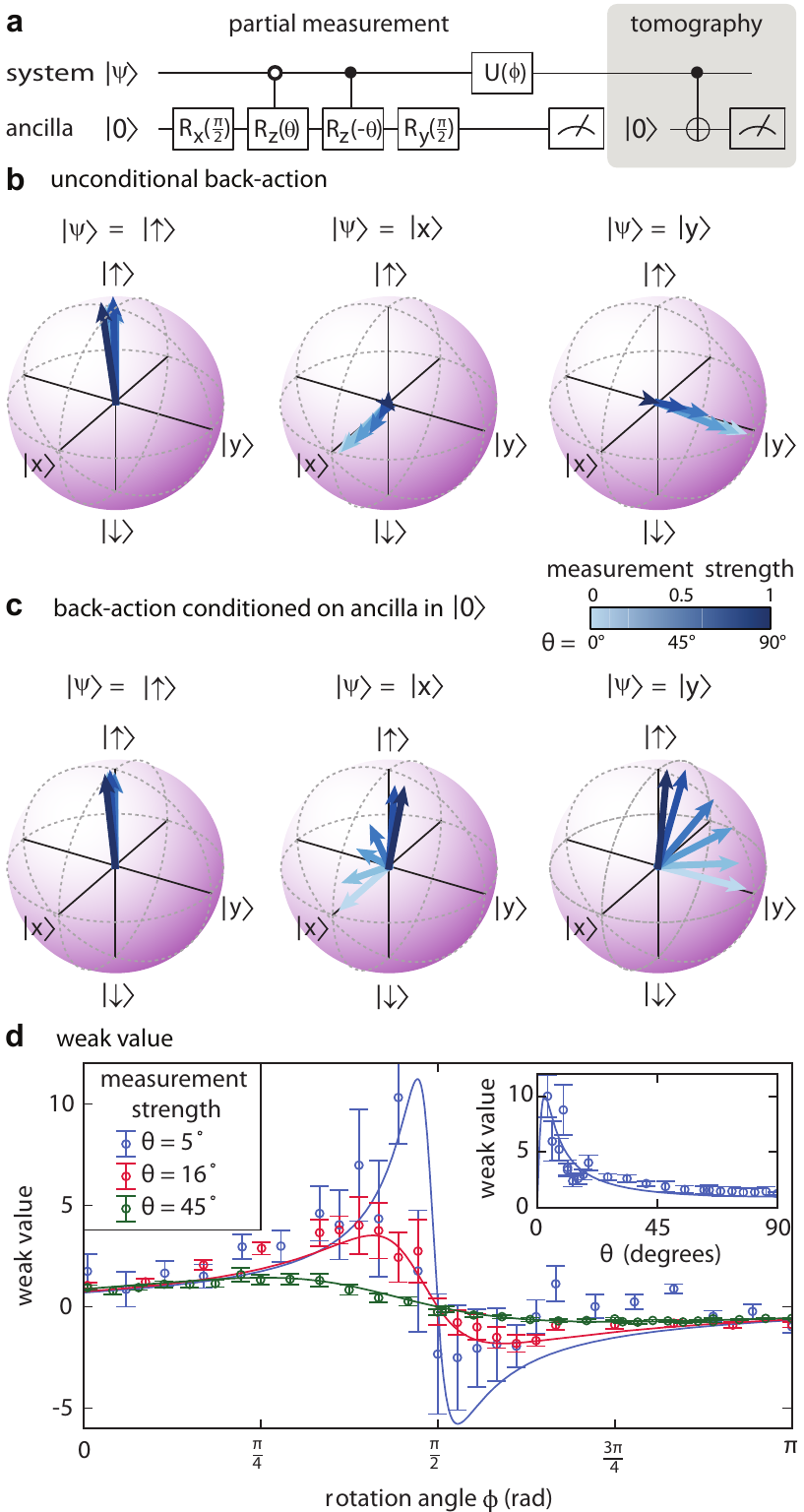}
	\caption{\label{fig2} Measurement backaction and quantum weak value. (a) We prepare an initial state  of the system ($\ket{\uparrow}$,  $\ket{x}$ and  $\ket{y}$), perform a partial measurement with strength $\theta$, and characterize the measurement backaction on the system by quantum state tomography. Quantum state tomography is implemented by an ancilla-assisted projective measurement, performed with the same protocol, setting $\tau = 229$ ns for $\theta = 90^{\circ}$. The nuclear spin basis rotation is performed with a $\frac{\pi}{2}$ radio-frequency pulse (along either $x$ or $y$). The basis rotation pulse for the tomography is applied before the readout of the ancilla, in order to avoid the dephasing induced by the state-characterization measurement (see main text). The data is corrected for errors in the readout and initialization of the system qubit, both of which are obtained from independent measurements. (b,c)  Measurement backaction for a partial measurement of increasing strength, independent of the measurement result for the ancilla qubit (b), or conditioned on the ancilla in  $\ket{0}$ (c). (d) Measurement of a modified weak value for the nuclear-spin qubit, performed by a partial measurement of strength $\theta$, followed by a strong measurement and post-selection of the state  $\ket{\downarrow}$, as a function of the basis rotation angle $\phi$ of the strong measurement. Inset: the modified weak value as a function of the strength $\theta$ of the partial measurement, setting the basis rotation angle of the strong measurement to the optimal value  $\phi = \frac{\pi}{2} - \theta$.}
\end{figure}

By combining a partial measurement with post-selection on the post-measurement state of the system, quantum weak values can be found that lie outside the range of eigenvalues of the measured observable~\cite{Brun_PhysRevA_2008,Groen_PRL_2013,Aharonov_PRL_1988,Pryde_PRL_2005,Dressel_ArXiv_2013}. A post-selected estimate of the nuclear spin in the $z$-basis gives $W = \frac{\bra{\psi_f} \hat{I}_z \ket{\psi_i}}{\bra{\psi_f} \psi_i \rangle}$ , where $ \psi_i (\psi_f )$ is the initial (final) state of the nucleus and $\hat{I}_z$ the Pauli $z$-operator reduced to a two-level system with eigenvalues +1 and $-$1. By post-selecting only on the final states having small overlap with the initial state, the measured value can be greatly amplified. Since our ancilla-based readout is digitized, we measure a modified weak value $W_m$~\cite{Groen_PRL_2013,Dressel_ArXiv_2013}, which reduces to $W$ in the limit of zero measurement strength $\theta = 0^{\circ}$. By sweeping the angle between initial and final state we observe up to 10-fold amplification ($W_m =10 \pm 3 $) (Fig.~2d), the highest reported for a solid-state system to date~\cite{Groen_PRL_2013}.

Using the partial measurements for measurement-based feedback requires reading out the ancilla without perturbing the system qubit. In our experiment the system qubit can dephase during ancilla readout both through a spin-flip of the electron in the course of optical excitation (Fig.~3b) and due to the difference in the effective nuclear g-factor in the electronic ground- and optically excited state~\cite{Jiang_PRL_2008}. Note that for the characterization of a single partial measurement (Fig.~2) we circumvent this dephasing by interchanging the measurement basis rotation and the ancilla readout; this interchange is not possible for real-time adaptive measurements.

To mitigate the nuclear dephasing during ancilla readout we reduce the ancilla spin-flip probability using a dynamical-stop readout technique. We partition the optical excitation time in short ($1~ \mu$s) intervals and we stop the excitation laser as soon as a photon is detected (Fig.~3a). This reduces redundant excitations without compromising the readout fidelity. In Fig. 3b we show the correspondence between pre- and post-measurement states for the two eigenstates of the ancilla. For the state $\ket{0}$ the dynamical-stop readout increases the fidelity ($F = \bra{\psi_i}\rho_m \ket{\psi_i}$) from 0.18 $\pm$ 0.02 to 0.86 $\pm$ 0.02. The latter fidelity is solely limited by the cases where the spin flipped before a photon was detected: we find $F = 1.00 \pm 0.02$ for the cases in which a photon was detected. As expected, the fidelity is high ($F = 0.996 \pm 0.006$) for input state $\ket{1}$ as this state is unaffected by the excitation laser. The dynamical-stop technique thus implements a quantum non-demolition (QND) measurement of the ancilla electron spin with an average fidelity of 0.93 $\pm$ 0.01 for the post-measurement state.

The dynamical-stop readout of the ancilla significantly reduces the dephasing of the nuclear spin qubit during measurement as shown in Fig.~3c. Starting with the nuclear spin in state $\ket{x} = \frac{\ket{0} + \ket{1}}{\sqrt{2}}$, a conventional readout of the ancilla completely dephases the nuclear spin in $\sim 25~\mu$s (the fidelity with respect to $\ket{x}$ is 0.5) while the fidelity of the dynamical-stop readout saturates to 0.615 $\pm$ 0.002 (likely limited by changes in the effective g-factor of the nuclear spin). The dynamical-stop readout thus leaves the system in a coherent post-measurement state that can be used in a real-time feedback protocol. 

Preserving coherence of the post-measurement state enables a proof-of-principle realization of measurement-only control, by implementing sequential measurements and tuning the strength of the second measurement in real time conditioned on the outcome of the first measurement (Fig.~4a). We choose as our target the creation of the state $\ket{\psi} = \cos{(\frac{\pi}{4}-\frac{\theta_1}{2})}\ket{\downarrow}+\cos{(\frac{\pi}{4}+\frac{\theta_1}{2})}\ket{\uparrow}$ from initial state $\ket{x}$ using only partial measurements of $\hat{I}_z$. The first measurement with strength $\theta_1$ will prepare either the desired state, or the state $\ket{\psi_{wrong}} =  \cos{(\frac{\pi}{4}+\frac{\theta_1}{2})}\ket{\downarrow}+\cos{(\frac{\pi}{4}-\frac{\theta_1}{2})}\ket{\uparrow}$ , each with probability 0.5. We adapt the strength of the second measurement $\theta_2$ according to the outcome of the first measurement: we set $\theta_2 = 0$ if the first measurement directly yielded the target state, but if the “wrong” outcome was obtained we set the measurements strength to

\begin{equation}
\theta_2 = \sin{^{-1}\left[2 \frac{\sin{\theta_1}}{1 + \sin{^2 \theta_1}}\right]},
\end{equation}

such that the second measurement will probabilistically rotate the qubit to the target state. The total success probability of this two-step protocol is  $p_{suc} = \frac{1}{2}(1 + \cos{\theta_1})$. In principle the protocol can be made fully deterministic~\cite{Ashhab_PhysRevA_2010} by incorporating a reset in the form of a projective measurement along the $x$-axis.

To find the improvement achieved by the feedback, we first compare the success probability of our adaptive measurement protocol to the success probability for a single measurement (Fig.~4b bottom panel). The success probability clearly increases with the adaptive protocol and is proportional to the readout fidelity of the $\ket{0}$ state of the ancilla, which is maximum for readout times  \textgreater~25~$\mu$s. The fidelity of the final state (Fig.~4b upper panel) is limited by the remaining dephasing of the system during readout of the ancilla as shown in Fig.~3b. This constitutes the trade-off between success probability and state fidelity. 

\begin{figure*}
	\includegraphics{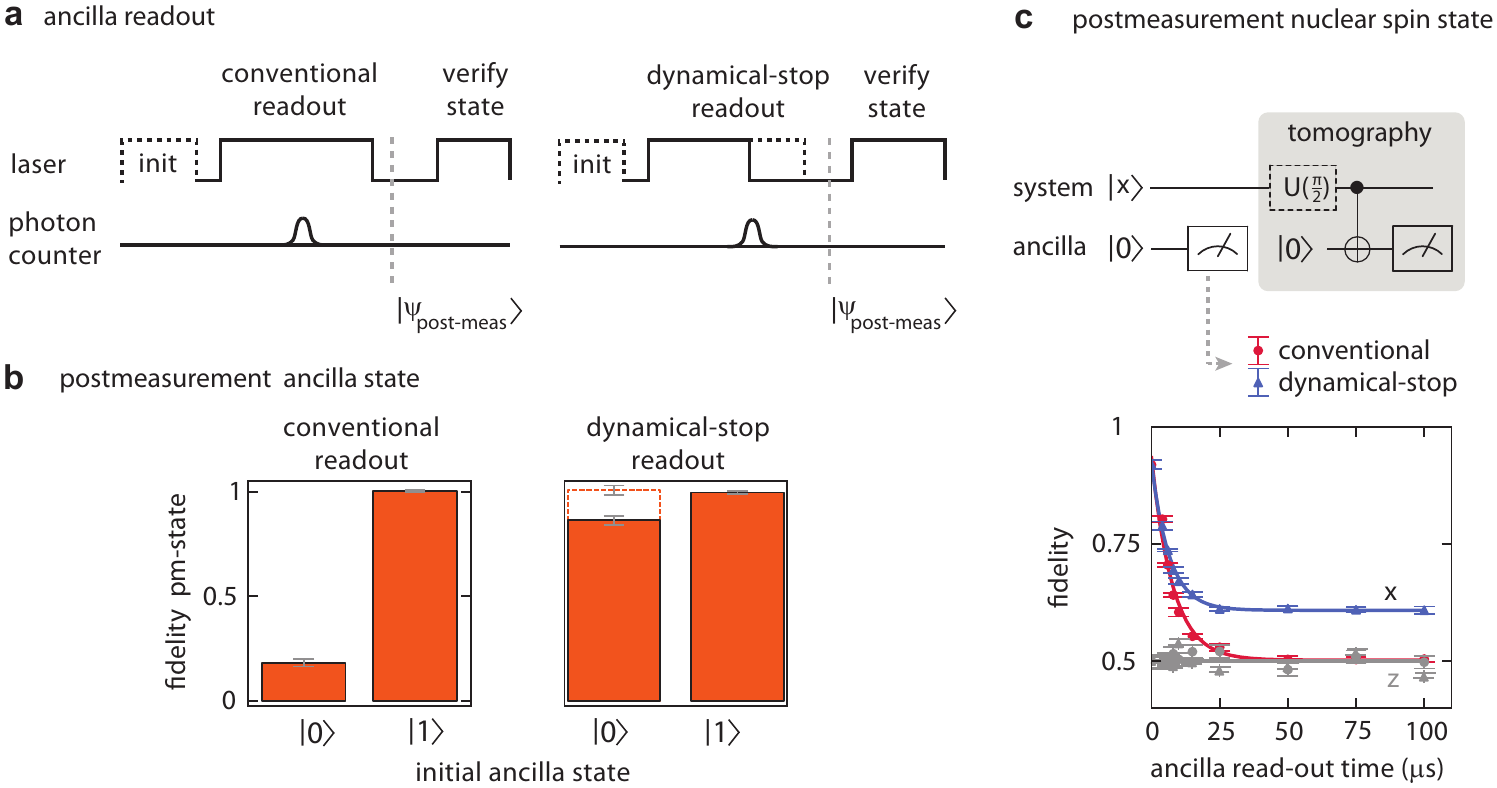}
	\caption{\label{fig3} Quantum non-demolition measurement of the ancila and system qubit coherence during readout. (a) The ancilla is initialized in $\ket{0}$ ($\ket{1}$) by optically pumping the $A_2$ ($E_y$) transition. The ancilla is then read out by exciting the $E_y$ transition for 100 $\mu$s (conventional readout), or until a photon was detected (dynamical-stop readout). Finally a conventional readout is performed to verify the post-measurement state. (b) Fidelity of the post-measurement state of the ancilla for conventional readout (left graph) and dynamical-stop readout (right graph). Results are corrected for the infidelity in the final readout.  (c) Coherence of the system qubit state after ancilla readout. The graph shows the fidelity of the system with respect to $\ket{x}$ for conventional readout (red) and dynamical-stop readout (blue). The $z$-component of the system is unaffected as shown by the constant fidelity with respect to $\ket{\downarrow}$ (grey). }
\end{figure*}

We show that the increase in success probability is non-trivial by comparing the final state fidelity with and without feedback (Fig.~4b upper panel). In principle the success probability can be increased in the absence of feedback by accepting a certain number of false measurement outcomes at the cost of a reduced fidelity. We calculate the maximum fidelity that can be achieved in this way by performing only the first measurement and increasing the success probability to that of the adaptive protocol using post-selection (grey line in Fig.~4b, upper panel). We find that the measured state fidelity in the adaptive protocol is above this bound (Fig.~4b, green area), which indicates that the adaptive measurement indeed successfully corrects the kickback from the first measurement, thus yielding a clear advantage over open-loop protocols.

We note that, in contrast to pioneering adaptive measurement experiments on photons that only used experimental runs in which a photon was detected at each measurement stage~\cite{Prevedel_Nature_2007}, our protocol is fully deterministic in the sense that the partial measurement always yields an answer. In particular, the data in Fig.~4 includes all experimental runs and thus no post-selection is performed, as desired for future applications in metrology and quantum computing. 

The performance of the protocol can be further improved by increasing the ancilla readout fidelity (either by improving the collection efficiency or reducing spin-flip probability) and by further reducing the dephasing of the system during readout. A particularly promising route is to use nuclear spins farther away from the NV center (e.g. Carbon-13 spins) that have much smaller hyperfine couplings~\cite{Zhao_NatureNano_2012,Taminiau_PRL_2012,Kolkowitz_PRL_2012} and are more robust against changes in the orbital state of the electron spin.

\begin{figure*}
	\includegraphics{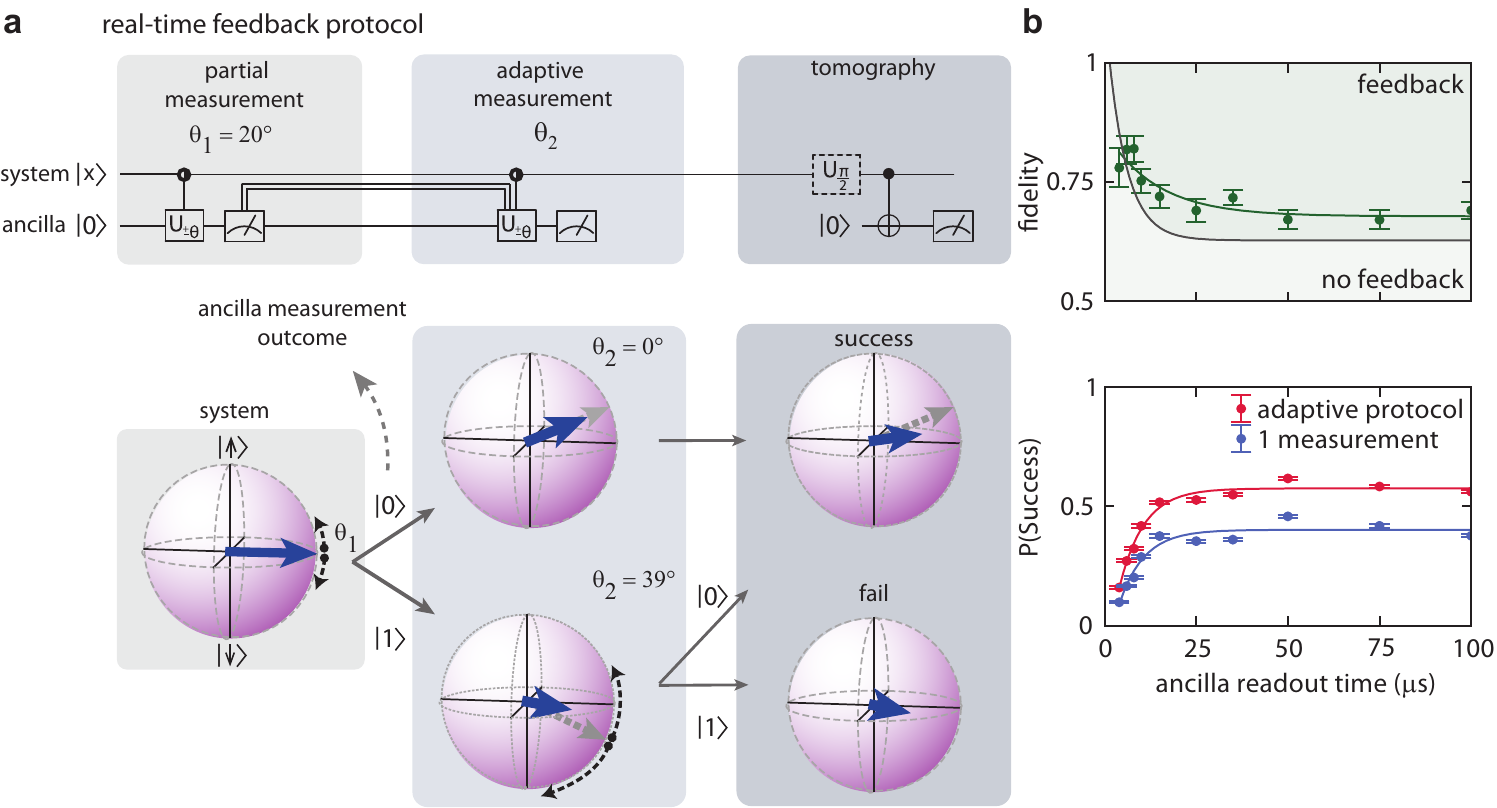}
	\caption{\label{fig4} Manipulation of a nuclear spin state by sequential partial adaptive measurements with real-time feedback. (a) Adaptive measurement protocol. The ancilla qubit is initialized in $\ket{0}$ and the system qubit is prepared in $\ket{x}$. The strength of the second measurement ($\theta_2$) is adjusted according to the outcome of the first measurement. The system is analysed by state tomography at each intermediate step. The result of the tomography is plotted on the bloch spheres (blue vector) and compared with the ideal case (grey vector). (b) Fidelity of the output state with respect to the target state as a function of ancilla readout time (dynamical-stop readout) with feedback (only the cases where the protocol heralds success). The grey line is obtained by performing one measurement and adding negative results to artificially increase the success probability to that of the adaptive protocol (red line in lower panel). In the lower panel we show the probability that the protocol heralds success for one measurement and for the adaptive protocol.  }
\end{figure*}

In conclusion, we implemented sequential partial measurements and showed that by adjusting the measurement strength in real-time we can steer a quantum system towards a desired state. Our work is the first experimental exploration of a fundamental concept in the field of quantum measurement and control~\cite{Wiseman_NatureNV_2011} that may find application in systems where control fields are difficult to generate. Furthermore, the use of adaptive measurements as presented here can increase the performance of spin-based magnetometers~\cite{Cappellaro_PhysRevA_2012,Higgins_Nature_2007}. Finally, our results can be combined with recently demonstrated methods for generating entanglement between separate NV center spins~\cite{Bernien_Nature_2013,Dolde_NatPhys_2013}. Taken together, these techniques form the core capability required for one-way quantum computing, where quantum algorithms are executed by sequential adaptive measurements on a large entangled ‘cluster’ state~\cite{Raussendorf_PRL_2001,Prevedel_Nature_2007}.

\section{Methods}
We use a naturally-occurring nitrogen-vacancy center in high-purity type IIa CVD diamond, with a \textless 111\textgreater-crystal orientation obtained by cleaving and polishing a \textless100\textgreater -substrate. Experiments are performed in a bath cryostat, at the temperature of 4.2~K, with an applied magnetic field of 17~G. Working at low-temperature, we can perform efficient electron spin initialization (F~=~0.983~$\pm$~0.006) and single-shot readout (the fidelity is 0.853~$\pm$~0.005 for $m_S = 0$ and 0.986~$\pm$~0.002 for $m_S = -1$) by spin-resolved optical excitation~\cite{Robledo_Nature_2011}. Initialization of the nuclear spin is done by measurement~\cite{Robledo_Nature_2011}, with fidelity  0.95~$\pm$~0.02. Single-qubit operations can be performed with high accuracy using microwave (for the electron) and radio-frequency (for the nucleus) pulses applied to the gold stripline. Note that the single-qubit operations on the nucleus are only used for state preparation and tomography, but not in the feedback protocol. The dephasing time $T_2^*$ is (7.8~$\pm$~0.2)~ms for the nuclear spin and (1.35~$\pm$~0.03)~$\mu$s for the electron spin. 

\section{Acknowledgements}
We thank L. DiCarlo, G. De Lange and L. Vandersypen for helpful discussions and comments, and R.N. Schouten and M.J. Tiggelman for technical assistance. We acknowledge support from the Dutch Organization for Fundamental Research on Matter (FOM), the DARPA QuASAR programme, the EU DIAMANT and S3NANO programmes and the European Research Council through a Starting Grant. Work at the Ames Laboratory was supported by the U.S. Department of Energy Basic Energy Sciences under contract no. DE- AC02-07CH11358.

\bibliographystyle{naturemag}
\bibliography{MBlok_adaptmsmnts_ref_papers}

\end{document}